\documentclass[aps,prl,twocolumn,superscriptaddress,groupedaddress]{revtex4-1}
\usepackage{graphicx}  
\usepackage{dcolumn}   
\usepackage{bm}        
\usepackage{amssymb}   
\usepackage{amsmath,amsthm} 
\usepackage{mathrsfs}
\begin{document}
\title{
Comment on ``Experimental demonstration of a universally valid error-disturbance uncertainty relation in spin measurements''
}
\author{Y. Kurihara}
\email[]{yoshimasa.kurihara@kek.jp}
\affiliation{
The High Energy Accelerator Organization $(KEK)$,Tsukuba, Ibaraki 305-0801, Japan}

\begin{abstract}
In this comment we show that the experimental results for a universally valid uncertainty relation in ref.\cite{citeulike:10228899} cannot be justified. The experiments cannot be recognized to establish a violation of the Heisenberg-type uncertainty relation suggested by Ozawa\cite{Masanao200321,PhysRevA.67.042105}.
\end{abstract}

\maketitle
\begin{figure}
\includegraphics[height=5cm]{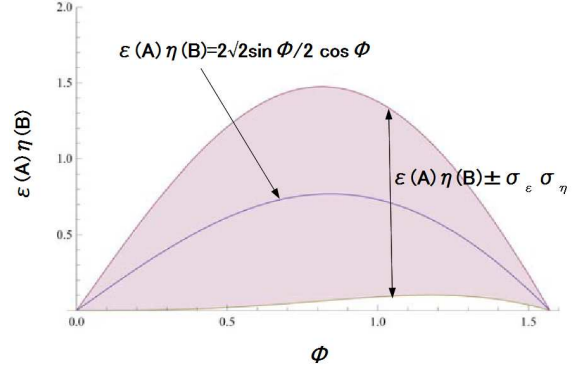}%
\caption{\label{fig1}The Heisenberg-pair of observables of ``error'' and ``disturbance'' defined in \cite{citeulike:10228899} as a function of a detuning angle of $\phi$. The simultaneous/successive measurements of two observables must have the Kennard-Robertson type lower bound of the uncertainty depicted by shadowed region centered the theoretical predictions of the observable.}
\end{figure}

Ozawa has proposed a universally valid uncertainty relation (UVUR hereafter) in refs.\cite{Masanao200321,PhysRevA.67.042105}. Moreover he has pointed out that there are possibilities to violate a standard Heisenberg uncertainty relation between a root-mean-square error and a disturbance. Before discussing the experimental results and their conclusions in detail, let us review his proof of the UVUR. Usually the UVUR refers such an Eq.(26) in \cite{Masanao200321} as
\begin{eqnarray*}
(26)~~~~~~
\epsilon(Q)\eta(P)+\epsilon(Q)\sigma(P)+\sigma(Q)\eta(P)\geq\frac{\hbar}{2},
\end{eqnarray*}
where $\epsilon(Q)$ is a root-mean-square error of an observable $Q$, $\sigma(Q)$ a standard deviation of $Q$, $\eta(P)$ a root-mean-square disturbance for an observable $P$, and $\sigma(P)$ a standard deviation of $P$. (Hereafter equation numbers refer ref.\cite{Masanao200321}) However he used Eqs.(29) to (31) for proving the UVUR of Eq.(26). Those equations are necessary conditions to establish the Eq.(26). Then the UVUR must refer as a set of equations form Eq.(16) to Eq.(31) with taking ``AND'' of all of them. Among them, especially Eq.(29)
\begin{eqnarray*}
(29)~~~~~~
\epsilon(Q)\eta(P)&\geq&\sigma(N(Q))\sigma(P(0))\\
&\geq&\frac{1}{2}|\langle[N(Q),D(P)]\rangle|
\end{eqnarray*}
is important, since it is treating the Heisenberg-pair. This condition is unavoidable restriction to discuss the violation of the Heisenberg-type uncertainty relation. Here $N(Q)$ is defined as 
\begin{eqnarray*}
(12)~~~~~~
N(Q)&=&M(\Delta t)-Q(0),
\end{eqnarray*}
where $Q(0)=Q\otimes I$ is an operator of the observable $Q$ on the system at $t=0$ (just before the measurement), and $M(\Delta t)=U^\dagger(I\otimes M)U$, where $U$ is an unitary operator for time evolution from $t=0$ to $t=\Delta t$ (just after the measurement), and $M$ an operator of the observable $P$ on the probe. Here $A\otimes B$ refers a direct product of the Hilbert space of the system $A$ and the probe $B$. 
$D(P)$ is defined as 
\begin{eqnarray*}
(4)~~~~~~
D(P)&=&P(\Delta t)-P(0),
\end{eqnarray*}
where $P(0)=P\otimes I$ and $P(\Delta t)=U^\dagger P(0)U$. The first inequality of Eq.(29) comes from the universal fact that the variance is not greater than the mean-square. The equality can hold if and only if the mean value of $N(Q)$ and $D(P)$ are zero.\\
 To advocate the violation of the Heisenberg-type uncertainty relation, $\epsilon(Q)=0$ or $\eta(P)=0$, one has to prove $\langle[N(Q),D(P)]\rangle=0$. In order to realize there are three possibilities to prepare the experimental condition which satisfy this condition.
\begin{enumerate}
\item $N(Q)$ and $D(P)$ are commutable:\\
It is impossible because $Q(0)$ is non-commutative  with $P(0)$ and $M(\Delta t)$ is non-commutative  with $P(\Delta t)$. Then operators $N(Q)$ and $D(P)$ cannot be commutable each other.
\item $N(Q)$ and/or $D(P)$ are $Null$-operators which eigenvalue is identically zero:\\
It is also impossible in general. For example, operators $M(\Delta t)$ and $Q(0)$ belongs to different Hilbert space, then if $N(Q)$ is $Null$-operator, both $M(\Delta t)$ and $Q(0)$ may be $Null$-operators. However that is contradict to a fact that $Q(0)$ has a finite variance. However there is one possibility to escape from this requirement that all of the eigenvalues of $M(\Delta t)$ and $Q(0)$ are completely the same. In ref.\cite{citeulike:10228899}, authors insist their experimental condition is prepared as this kind of situation. It will be discussed later. 
\item eigenvalues of $N(Q)$ and/or $D(P)$ are zero during measurement $(0\leq t \leq\Delta t)$:\\
These condition can be kept during the typical time period which determined by the inverse of the typical energy scale of the process. That is, for example, order of $10^{-17}{\rm s}$ for the atomic energy scale, and order of the Planck time for the measurement of gravitational phenomena. It is very difficult to terminate the interaction after this time period, or to keep interaction during this period, depend on a strength of the coupling.
\end{enumerate}
 ~Now let us investigate the experimental condition of ref.\cite{citeulike:10228899} in detail. A theoretical background of this experiment is described in ref.\cite{ozawa-2005-7}. The experimental set-up follows the example described in section~5 in ref.\cite{ozawa-2005-7}. (Hereafter, section and equation numbers refer ref.\cite{ozawa-2005-7} unless otherwise stated.) In section~5, Ozawa shows that the ``{\it projective measurement}'' can give a perfect matching between eigenvalues of $M(\Delta t)$ and $Q(0)$. Terminology in ref.\cite{ozawa-2005-7}, those are
\begin{eqnarray*}
(19)~~~~~~
N_a=U^\dagger(I\otimes M)U-A\otimes I,
\end{eqnarray*}
where $M(\Delta t)$ corresponds to $U^\dagger(I\otimes M)U$ and  $Q(0)$ to $A\otimes I$. In the experiment, that is realized as three non-commutative measurements of a neutron spin for three independent spacial directions. However, as known well, the measurement of the spin-state to specific direction cannot be described using any unitary operator of the time evolution as Ozawa described as ``{\it projective}''. That means the operator 
\begin{eqnarray*}
(49)~~~~~~
\Pi_m=\langle\xi|U^\dagger(I\otimes E_m^M)U|\xi\rangle_{\cal K}
\end{eqnarray*}
is NOT well defined. This unitary operator must be exist in order to evolute back the operator from $t=\Delta t$, after the measurements of $M1$ and $M2$, to $t=0$ just before the measurements. Otherwise one cannot use the equal-time commutation relation at $t=0$. Eq.(76)
\begin{eqnarray*}
(76)~~~~~~
\sum_m mM_m=A
\end{eqnarray*}
cannot be justified at $t=0$ because Eq.(49) cannot be held without the unitary operator $U$ and its inverse $U^\dagger$. Of course if whole system including systems and detectors is considered, there are some kind of unitary operator which represent time evolution due to an overall probability conservation, however that unitary operator cannot evolute back the ``projected system'' to that before the measurement. Then this kind of existence theorem cannot justify the existence of eq.(76).If one try to evolute back the probe operator $M$ from after the $\sigma_y$ measurement to before the $\sigma_x$ measurement, by brute force way, one cannot say the neutron spin direction is $|+z\rangle$, but can be any direction because of the lack of the unitary operator. Then one have to consider that $\epsilon(A)=1$ where the Heisenberg-type uncertainty relation is still held.\\
\indent
Even if the existence of the unitary operator is care nothing for the discussion, there are another question for the experiment.
The purpose of this experiment is to measure the Heisenberg-pair of $\epsilon(A)\eta(B)$. Here let us consider the Kennard-Robertson type lower bound of uncertainty for this observable. From the operator definition of $\epsilon(A)$ and $\eta(B)$ given in ref.\cite{citeulike:10228899}, 
a lower bound of the Kennard-Robertson type uncertainty relation can be given as
\begin{eqnarray*}
\sigma_\epsilon\sigma_\eta&=&\frac{1}{2}
\| 
[(\sigma_\phi-\sigma_x),\sqrt{2}[\sigma_\phi,\sigma_y]]|\psi\rangle
\|,\\
&=&\sqrt{\cos{\phi}\sin{\phi}\sin{2\phi}},
\end{eqnarray*}
which is drawn in FIG.\ref{fig1}. All terminologies are defined as the same as in ref.\cite{citeulike:10228899}. 
This uncertainty cannot be avoided if the experiment were truly the simultaneous or successive measurements of the uncommutative pair of observables. However this error band is obviously wider than errors assigned in {\bf Figure~5} in ref.\cite{citeulike:10228899}, which means the measurements are {\it NOT} the simultaneous/successive. As described in {\bf Method} section of the paper, measurements of the ``error'' and ``disturbance'' have been done with combination of four independent measurements with initial stats of
$
|+z\rangle,~|-z\rangle,~|+x\rangle,
$
and
$|+y\rangle$. An actual observable and uncertainty of $\epsilon(A)\eta(B)$ performed in the experiment can be describes more exactly as
\begin{eqnarray*}
\epsilon(A)\eta(B)&=&2\sqrt{2}\sin{\frac{\phi_1}{2}}\cos{\phi_2}
|_{\phi_{1,2}\rightarrow\phi},\\
\sigma_\epsilon\sigma_\eta&=&
\sqrt{2}\sin{\phi_1}\cos{\phi_2}.
\end{eqnarray*}
In this case, the uncertainty for the observable from four independent measurements must be estimated with averaged out for unmeasured parameters as 
\begin{eqnarray*}
\bar\sigma_\epsilon\bar\sigma_\eta&=&
\left\{
\int_{-\frac{\pi}{2}}^{0}d\phi_1+\int_{\frac{\pi}{2}}^{\pi}d\phi_1
+\int_{0}^{\frac{\pi}{2}}d\phi_2+\int_\pi^{\frac{3\pi}{2}}d\phi_2
\right\}
\sigma_\epsilon\sigma_\eta\\
&=&0.
\end{eqnarray*}
That is the reason why the experiment can measure the Heisenberg-pair of observables $\epsilon(A)\eta(B)$ as precise as shown in the paper without disturbed by the Kennard-Robertson type uncertainty. At the same time, this analysis shows the measurements in ref.\cite{citeulike:10228899} cannot be recognized as the simultaneous/successive measurements.
\\
\indent
In conclusion, the experimental set-up of ref.\cite{citeulike:10228899} is not satisfied the necessary condition to be true for the UVUR proved by Ozawa in \cite{Masanao200321,PhysRevA.67.042105}. It must be stressed that the UVUR refers as a set of equations from Eq.(16) to Eq.(31) in ref.\cite{Masanao200321} with taking ``AND'' of all of them.
\bibliography{comment}

\begin{thebibliography}{4}%
\makeatletter
\providecommand \@ifxundefined [1]{%
 \@ifx{#1\undefined}
}%
\providecommand \@ifnum [1]{%
 \ifnum #1\expandafter \@firstoftwo
 \else \expandafter \@secondoftwo
 \fi
}%
\providecommand \@ifx [1]{%
 \ifx #1\expandafter \@firstoftwo
 \else \expandafter \@secondoftwo
 \fi
}%
\providecommand \natexlab [1]{#1}%
\providecommand \enquote  [1]{``#1''}%
\providecommand \bibnamefont  [1]{#1}%
\providecommand \bibfnamefont [1]{#1}%
\providecommand \citenamefont [1]{#1}%
\providecommand \href@noop [0]{\@secondoftwo}%
\providecommand \href [0]{\begingroup \@sanitize@url \@href}%
\providecommand \@href[1]{\@@startlink{#1}\@@href}%
\providecommand \@@href[1]{\endgroup#1\@@endlink}%
\providecommand \@sanitize@url [0]{\catcode `\\12\catcode `\$12\catcode
  `\&12\catcode `\#12\catcode `\^12\catcode `\_12\catcode `\%12\relax}%
\providecommand \@@startlink[1]{}%
\providecommand \@@endlink[0]{}%
\providecommand \url  [0]{\begingroup\@sanitize@url \@url }%
\providecommand \@url [1]{\endgroup\@href {#1}{\urlprefix }}%
\providecommand \urlprefix  [0]{URL }%
\providecommand \Eprint [0]{\href }%
\providecommand \doibase [0]{http://dx.doi.org/}%
\providecommand \selectlanguage [0]{\@gobble}%
\providecommand \bibinfo  [0]{\@secondoftwo}%
\providecommand \bibfield  [0]{\@secondoftwo}%
\providecommand \translation [1]{[#1]}%
\providecommand \BibitemOpen [0]{}%
\providecommand \bibitemStop [0]{}%
\providecommand \bibitemNoStop [0]{.\EOS\space}%
\providecommand \EOS [0]{\spacefactor3000\relax}%
\providecommand \BibitemShut  [1]{\csname bibitem#1\endcsname}%
\let\auto@bib@innerbib\@empty
\bibitem [{\citenamefont {Erhart}\ \emph {et~al.}(2012)\citenamefont {Erhart},
  \citenamefont {Sponar}, \citenamefont {Sulyok}, \citenamefont {Badurek},
  \citenamefont {Ozawa},\ and\ \citenamefont {Hasegawa}}]{citeulike:10228899}%
  \BibitemOpen
  \bibfield  {author} {\bibinfo {author} {\bibfnamefont {J.}~\bibnamefont
  {Erhart}}, \bibinfo {author} {\bibfnamefont {S.}~\bibnamefont {Sponar}},
  \bibinfo {author} {\bibfnamefont {G.}~\bibnamefont {Sulyok}}, \bibinfo
  {author} {\bibfnamefont {G.}~\bibnamefont {Badurek}}, \bibinfo {author}
  {\bibfnamefont {M.}~\bibnamefont {Ozawa}}, \ and\ \bibinfo {author}
  {\bibfnamefont {Y.}~\bibnamefont {Hasegawa}},\ }\href {\doibase
  10.1038/nphys2194} {\bibfield  {journal} {\bibinfo  {journal} {Nat Phys}\
  }\textbf {\bibinfo {volume} {advance online publication}} (\bibinfo {year}
  {2012}),\ 10.1038/nphys2194}\BibitemShut {NoStop}%
\bibitem [{\citenamefont {Ozawa}(2003{\natexlab{a}})}]{Masanao200321}%
  \BibitemOpen
  \bibfield  {author} {\bibinfo {author} {\bibfnamefont {M.}~\bibnamefont
  {Ozawa}},\ }\href {\doibase 10.1016/j.physleta.2003.07.025} {\bibfield
  {journal} {\bibinfo  {journal} {Physics Letters A}\ }\textbf {\bibinfo
  {volume} {318}},\ \bibinfo {pages} {21 } (\bibinfo {year}
  {2003}{\natexlab{a}})}\BibitemShut {NoStop}%
\bibitem [{\citenamefont {Ozawa}(2003{\natexlab{b}})}]{PhysRevA.67.042105}%
  \BibitemOpen
  \bibfield  {author} {\bibinfo {author} {\bibfnamefont {M.}~\bibnamefont
  {Ozawa}},\ }\href {\doibase 10.1103/PhysRevA.67.042105} {\bibfield  {journal}
  {\bibinfo  {journal} {Phys. Rev. A}\ }\textbf {\bibinfo {volume} {67}},\
  \bibinfo {pages} {042105} (\bibinfo {year} {2003}{\natexlab{b}})}\BibitemShut
  {NoStop}%
\bibitem [{\citenamefont {Ozawa}(2005)}]{ozawa-2005-7}%
  \BibitemOpen
  \bibfield  {author} {\bibinfo {author} {\bibfnamefont {M.}~\bibnamefont
  {Ozawa}},\ }\href {doi:10.1088/1464-4266/7/12/033} {\bibfield  {journal}
  {\bibinfo  {journal} {QUANTUM SEMICLASS.OPT.}\ }\textbf {\bibinfo {volume}
  {7}},\ \bibinfo {pages} {S672} (\bibinfo {year} {2005})}\BibitemShut
  {NoStop}%
\end{thebibliography}%

\end{document}